\documentstyle[aps,pre,epsfig,multicol]{revtex}
\begin{document}
\draft
\title{On the universality of distribution of ranked 
cluster masses at critical percolation}

\author{Parongama Sen}

\address{Department of Physics, Surendranath College, 24/2 Mahatma Gandhi Road,
Calcutta 700009}

\maketitle
\begin{abstract}
The distribution of  masses of clusters smaller than the infinite cluster 
 is evaluated at the percolation threshold. 
The clusters are ranked according to their masses  and the distribution 
$P(M/L^D,r)$ of the scaled masses $M$ for any 
rank $r$ shows a universal behaviour for different lattice sizes $L$ ($D$ is
the fractal dimension).
For different ranks however, there is a universal distribution function only
in the large rank limit, i.e., 
$P(M/L^D,r)r^{-y\zeta } \sim g(Mr^y/L^D)$ ($y$ and $\zeta$ are defined in the 
text), 
where the universal scaling function $g$ is found to be Gaussian in nature.
\end{abstract}  

\begin{multicols}{2}

Percolation is a classic example of  systems with quenched disorder \cite{SA}. In a 
discrete lattice, sites or bonds are present with a certain probability
and clusters are formed by connecting neighbouring occupied sites.
At a critical probability, an "infinite" cluster appears for the first time
which spans the whole lattice.   

The average mass or size of the spanning cluster is known to scale as 
$M \sim  L^D$, where $L$ is the lattice size and $D$  the fractal dimension.
In two  recent papers \cite{wata,JSA}, it was shown that when the clusters
are ranked, the average masses of the ranked  clusters also
show a similar scaling behaviour. This is true  even for the clusters of
large ranks,  
 which 
are definitely smaller than the spanning cluster. These clusters  
  have been 
termed  "effectively spanning" as their masses diverge with the
lattice size although they do not really span the lattice. 
The behaviour of the average scaled mass $M/L^D$ as a function of the rank $r$
was found to be  
\begin{equation}
\label{eq1}
\langle M/L^D \rangle \sim r^{-\lambda},
\end{equation}
where $\lambda$ can be expressed in terms
of  other
known exponents of percolation as \cite{JSA} 
\begin{equation}
\label{eq2}
\lambda = 1/(\tau -1).
\end{equation}
Here $\tau = 1+ d/D$ where $d$ is the spatial dimension.
It was also argued that the above behaviour is observed  only in the asymptotic 
limit $r \rightarrow \infty$. The  $\langle M/L^D\rangle $ vs. $r$ curve 
actually changes its slope
slowly (in a log-log plot). Hence for a given range of $r$, one can define an effective 
$\lambda_{eff}(r)$ 
with
$\lambda_{eff}(r \rightarrow \infty) $ given by (\ref{eq2}).
Very large rank would mean essentially clusters of size
one or two in a finite lattice and these are not of present interest.

Distribution function and its moments are useful 
for studying important properties of a system like 
multifractality, lacunarity etc.
Distribution of 
the  size of  {\it all} the clusters, which is essentially
 the number of clusters of a given size (as a function of the 
size)  in a dilute lattice, 
is well-known \cite{SA}  both  
at and away from
 criticality. 
Distributions of
several other quantities like the size of the spanning clusters,
chemical distances, shortest and longest paths on the percolation cluster etc. have 
also been studied in detail  \cite{Hav1,Hov,PS1,PS2,porto,grass}. 
In general, at criticality, when the randomness is relevant, a universal
non-Gaussian distribution function  will exist \cite{Aha2}.
Although questions about distributions functions have been addressed
for quite some time,  still a  proper understanding is lacking 
in several areas \cite{Hav1,PS2,porto,grass,ziff}.
Recently, the behaviour of the distribution of the largest
cluster below criticality was also studied \cite{Mar}.

The existence  of universal scaling functions for
the different quantities in the percolating lattice  and the properties of the
ranked clusters inspired us  to study  the
distribution of the mass or size of these clusters at the percolation
threshold. Although most of the quantities which
have been studied earlier are directly related to the 
percolating or spanning cluster (like the mass of the percolating cluster,
the mass of the backbone, shortest path on the backbone etc.)
the smaller clusters are no less important.  In addition the 
remarkable fractal-like behaviour of the ranked clusters calls
for further investigations. 
Our interest is particularly on the question  of 
universality of the distribution function. 

In the simulation, the clusters in a square
lattice (with helical boundary condition) are identified 
using the Hoshen-Kopelman algorithm.
We rank the clusters  at the 
percolation threshold irrespective of the fact whether the 
lattice is actually percolating or not.
It may be noted that the ranked clusters may have  degeneracy
in the sense that there may be several clusters with
the same rank in a particular realisation of the lattice. 
We checked, however, that incorporating this degeneracy hardly
affects the results. 

We first check the fact that the 
  slope 
of the average cluster mass $\langle M/L^D \rangle$ vs.  $r$ in a log
log plot is indeed not unique in spite of (\ref{eq1}) and in agreement with 
\cite{JSA}.
  We also verify that
$\lambda_{eff} (r) $ has very weak finite size dependence, if any, 
as shown in Fig. 1. For $r > 30$, there is apparently some
size dependence, but for the small lattices (e.g. $L = 200,300$ etc.), 
 such ranks correspond to clusters which are not effectively
spanning. Indeed,  in \cite{JSA}, the asymptotic value of  $\lambda$ 
was found from very large lattices. 
However, one can obtain useful information as long as distribution functions
are concerned even from relatively smaller lattices.

The number of clusters of rank $r$ with mass $M/L^D$ is evaluated.
The normalised probability distribution 
 ($P(M/L^D,r)$) 
of a cluster of scaled mass $M/L^D$ and rank $r$ 
 is obtained by dividing this number by the total 
number of clusters of rank $r$. 
This is shown for the ranks 4,6,10 and 14 for several lattice 
sizes in Fig. 2.
As in \cite{PS2}, where distributions for the case $r = 1$ 
were  considered only, the bin sizes are proportional to
$1/L^D$, and one directly obtains a universal distribution for
$P(M/L^D,r)$ for several values of $L$.
Another interesting feature is, as one plots $P(x = M/L^D,r)$ for several ranks, it is found that
the peaks of the distribution functions behave as 
$P_{max}(x_{max}) = x_{max}^{-\zeta}$ 
where  $x_{max}$ is the value of $x$ at which the peak occurs. 
(This is shown by the straight line touching the peaks 
of the distribution in Fig. 2. in a log-log plot.)
This   behaviour of the peaks persist  with a rank-independent 
value of $\zeta \simeq 1.25$ even for the higher ranks.

We are more interested, however,  in the behaviour of the 
probability distribution functions for different ranks for the
same lattice size. 
 The peak of the distribution  $P(M/L^D,r)$  
has a functional dependence  
on $r$   as $r^{\zeta \lambda_{eff}(r)}$ from the above 
mentioned behaviour
 and eq (\ref{eq1}).
However, in general the behaviour of the entire distribution
may not be as $r^{\zeta \lambda_{eff}(r)}$ and we observe that it is 
better to expect a general form as 

\begin{equation}
\label{scaling}
P(M/L^D,r)r^{-y\zeta } \sim g(Mr^y/L^D) 
\end{equation}
when plotted
against the natural scaling argument $Mr^y/L^D$.
Here  $y$ is   expected to 
be close to $\lambda_{eff}(r)$.  The values of $y$ are compared to 
different values of $\lambda_{eff}(r)$
(corresponding to  three different ranges  of $r$)   where the latter are 
obtained from a piecewise least 
square fitting of $\langle M/L^D\rangle $ vs $r$ curves. 
$g$ is a universal scaling function. Our attempt is to 
check whether one can actually obtain such a universal function
for the distributions. 

While the data for the smaller ranges of $r$ are taken from 
a system of $L = 500$, those   in Fig. 5  correspond to 
that with $L = 1000$. The number of random configurations generated 
are $10^4$ and $10^3$ respectively for the two sizes.
The results for the different ranges of $r$ (as appropriate to 
the system sizes considered) 
are summarised below:

{\it Small r}:  For $ 4 <  r < 14$, we find that only one part of the curves
(that beyond the peak value of $P(x,r)$) are collapsing when plotted against 
the proposed scaling 
argument with $y \simeq 1.75$. Here the actual value of $\lambda_{eff}(r)$ 
is around $1.45$. 

{\it Intermediate r}: For $r$ values in a higher range ($ 24 < r < 40$), we find that the two 
parts of the curves collapse separately with different values of $y$; 
$y  \simeq  1.4$ for the portion beyond the peak, and $y  \simeq  0.95 $ for the other 
portion. 
The value of $\lambda_{eff}$ for this range of $r$ is found
to be close to 1.25.

{\it Higher r}: When one plots the scaled probabilities for even higher values of  
$r$, we find for the first time, a simultaneous collapse of both  
sides of the curves with $y$ between 1 and 1.1.  
The value of $\lambda_{eff} $ in this range is also $\simeq 1.1$
Hence a universal function is indeed obtained for large $r$ values.
We believe, for even higher ranges of $r$ (for which reliable  data can be  obtained from  larger
lattices), the same behaviour will persist, with the
value of $y$ approaching the asymptotic value of $\lambda$. 
Interestingly, for the smaller and higher ranges of $r$, $y$ is neither equal to
$\lambda_{eff}$ or the asymptotic value of $\lambda $ (at least for
large $x$). However, in the scaling regime (i.e., for the 
large ranks), $y \simeq \lambda_{eff}$.

The major portion of the universal distribution seems to fit well with
a Gaussian distribution function of the form $\exp(-(x-x_0)^2/\sigma )$ 
with $0 < x < \infty$,  $\sigma \simeq 0.0005$, $A \simeq  0.17$ and $x_0 \simeq 0.1$.

Hence we obtain a distribution function in the following form

\begin{equation}
\label{gauss}
P(M/L^D,r) \sim r^{y \zeta} \exp (-(Mr^y/L^D -0.1)^2)/\sigma) 
\end{equation}
with $y \simeq 1.1$ and $\zeta \simeq 1.25$ for the higher ranks.

Hence we obtain two  most significant results in the present study:

a) The exponent  $\zeta \simeq  1.25$  for all ranges of
the ranks. This is significant as while other properties 
of the system are rank dependent, this particular one
remains constant.

b) The existence of a Gaussian distribution:
 Most of the distribution functions studied earlier
have yielded a more complicated universal function \cite{Hav1,Hov,PS2}. 
However, here also  the data corresponding to very small 
values of $Mr^y /L^D$ do not fall on the Gaussian fitting
curve.

It is difficult
to relate $\zeta$ to the known exponents in percolation. Naively,
if (\ref{eq1}) is to be derived from (\ref{gauss}), then
\begin{equation}
\langle M/L^D\rangle = \int \frac {M} {L^D} P(M/L^D,r) d(\frac {M}{L^D})
\sim r^{-\lambda}
\end{equation}
gives $\zeta = 1.0$ with $y = \lambda$. This involves the approximation that the 
mass of the cluster varies from zero to infinity.  This 
approximation and also possible deviations from the Gaussian distribution 
may be responsible for 
the discrepancy between this  value and the obtained value of $\zeta$; or it
may simply be due to errors in numerical estimate.

As already mentioned, the  distribution for the probability (per site)
of clusters with $s$ sites is known to be  $Q(s) \sim  s^{-\tau}$ in a 
percolating lattice.  
One may expect that this  behaviour can  be
extracted from $P(M/L^D,r)$ by calculating   
$ \Sigma_r  P(M/L^D,r) $
as  $s = (M/L^D) L^{D-d}$,  and a theoretical estimate of $\zeta$ can be made.
However, it has numerically been verified that one needs to
include clusters of all ranks to obtain $Q(s)$  in the above manner, 
and  the absence of a universal scaling law for all $r$
thus does not allow one to theoretically estimate $\zeta$.
 
%
In conclusion, the fact that the average  
cluster size approached a rank independent 
scaling form given by (\ref{gauss}) only for large $r$, is consistent 
with our result that the universal form is obtained again only in the 
large $r$ limit. One needs an exponent $y$ to obtain
a collapse of the data which should apparently equal $\lambda $. However, 
$y$ is greater than $\lambda_{eff}$ for the lower
ranges of the rank. Surprisingly though, for the 
intermediate range of the rank, a data collapse is achieved for 
the smaller masses with a value of $y$ very close to the
asymptotic value of $\lambda$. It is not clear  how
significant is this equivalence and  whether it is purely accidental.
In addition, we get  an exponent $\zeta$ from the
scaling behaviour of the probability distribution which is independent of the
rank.
An  approximate estimate of $\zeta$ is attempted to compare with
the numerically obtained value. 
As in the cases of other quantities in percolation, here also 
a   universal function is seen to exist, which in contrast to the others
 is  a simple Gaussian. The universal functions existing for each rank separately 
for several system sizes, however,  have  more complicated nature.

The author is grateful to the computing centre of Institute of Physics, Bhubaneswar,
where the programmes were run on a HPK9000/879 machine.
She also thanks D. Stauffer for a critical reading of the manuscript and very useful comments.

\narrowtext
\begin{figure}
\psfig{file=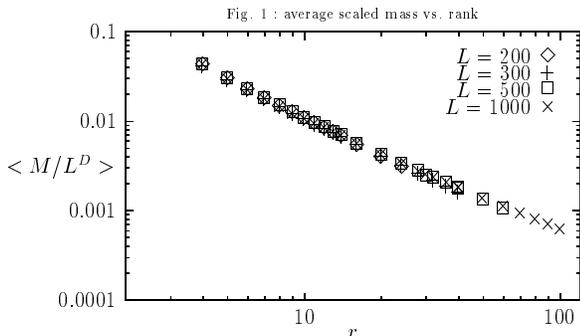,width=3in}
\caption{Variation of the average $\langle M/L^D\rangle $ against the ranks are shown for lattice sizes $L = 200,300,500$ and 1000. Larger number of
data (corresponding to higher ranks) are available for
increasing lattice size.}
\end{figure}

\begin{figure}
\psfig{file=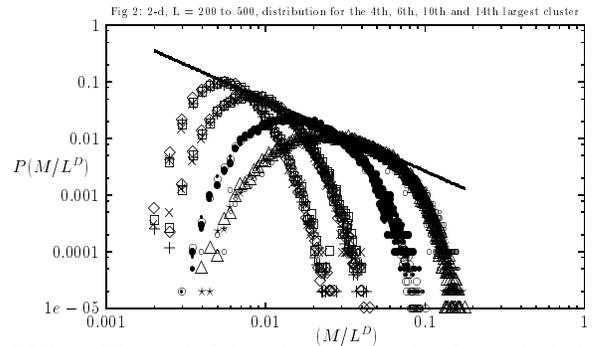,width=3in}
\caption{The probability distribution for the ranked cluster masses 
are shown for ranks $r = 4,6,10 $ and 14 (from right to left) 
for lattice sizes 
$L = 200,300,400$ and 500 against the scaled masses. 
 The peaks of the distribution show a power law behaviour with
$M/L^D$ where the maxima occur for each rank.}
\end{figure}

\begin{figure}
\psfig{file=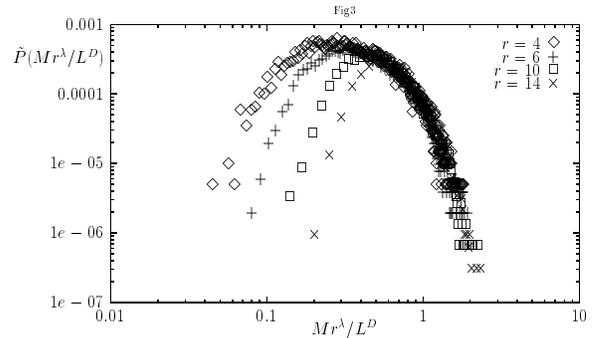,width=3in}
\caption{The partial collapse of the data for  the
scaled distribution $\tilde P(Mr^y/L^D) = P(M/L^D,r)r^{-y\zeta}$ 
is shown for ranks $r = 4,6,10$ and 14.
The value of $y$ is 1.75 here.} 
\end{figure}

\begin{figure}
\psfig{file=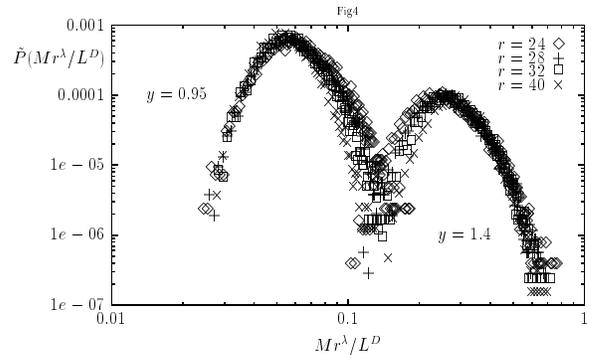,width=3in}
\caption{The partial collapses of the data for the scaled distribution  
 $~~~\tilde P(Mr^y/L^D)  =  P(M/L^D)r^{-y\zeta}~~$ 
for ranks $r = 24,28,32$ and $40$ with two different values of $y$ are shown separately.} 
\end{figure}

\begin{figure}
\psfig{file=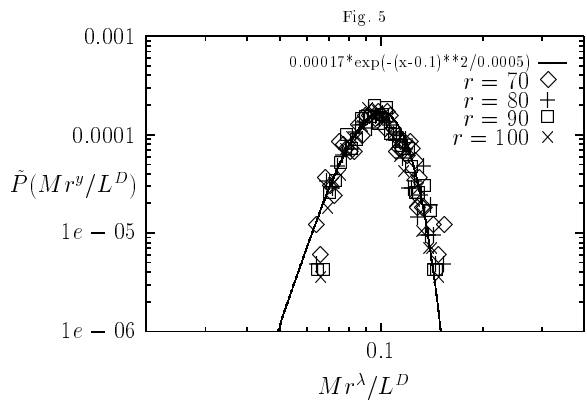,width=3in}
\caption{The  collapse  of the data for 
the scaled distribution 
 $\tilde P(Mr^y/L^D)  
 =  P(M/L^D,r)r^{-y\zeta}$ 
is shown for the ranks $r = 70,80,90 $ and 100.
The value of $y$ is 1.1 here.} 
\end{figure}

\end{multicols}


\begin{references}
\bibitem{SA} D. Stauffer and A. Aharony, {\it Introduction to
Percolation Theory} (Taylor and Francis, London, 1994).
\bibitem{wata} M. S. Watanabe, Phys. Rev. E {\bf 53} 4187 (1996). 
\bibitem{JSA}
N. Jan,  D. Stauffer and A. Aharony, J. Stat. Phys. {\bf 92} 325 (1998).
\bibitem{Hav1}
S. Havlin, B. L. Trus, G. H. Weiss and D. Ben-Avraham,  J. Phys A {\bf 18} L247 (1985). 
A. U. Neumann and S. Havlin J Stat Phys {\bf 52} 203 (1988)  
 H. Saleur and B. Derrida, J. Physique {\bf 46} 1043 (1985).  
\bibitem{Hov} J.-P. Hovi and A. Aharony,  Phys. Rev {\bf 56} 172 (1997).
\bibitem{PS1} P. Sen and P. Ray,  J Stat Phys {\bf 59} 1573 (1990). 
\bibitem{PS2} P. Sen, 
Int.J. Mod. Phys. C {\bf 10} 747 (1999).   
\bibitem{porto} M. Porto, S. Havlin, H. E. Roman and A. Bunde, Phys. Rev. E {\bf 58} R5205 (1998).  
\bibitem{grass} P. Grassberger, cond-mat/9906309.
\bibitem{Aha2}
A. Aharony and A. B. Harris, Phys. Rev. Lett. {\bf 77} 3700 (1996).
\bibitem{ziff} R. Ziff, cond-mat/9907305.
\bibitem{Mar} M. Z. Bazant, cond-mat/9905191.
\end{references}
\end{document}